\documentclass[11pt]{article}
\pdfoutput=1

\usepackage{titling}
\usepackage{amsmath}
\usepackage{slashed}
\usepackage{amssymb}
\usepackage{epsfig}
\usepackage{graphicx}
\usepackage{multirow}
\usepackage{color}
\usepackage[normal]{subfigure}
\usepackage{rotating}
\usepackage{hyperref}
\usepackage[margin=0.9in]{geometry}
\usepackage[table]{xcolor}
\usepackage[utf8x]{inputenc}
\usepackage[compress,numbers,sort]{natbib}
\usepackage{authblk}
\usepackage{colortbl}
\usepackage{pdflscape}
\usepackage{color}
\usepackage{mathtools}
\usepackage{tabu}
\usepackage{braket}
\usepackage{arydshln}
\usepackage{enumerate}
\usepackage{placeins}

\newcommand{\published}[1]{%
\gdef\puB{#1}}
\newcommand{\puB}{}

\title{Exploring the flavour structure of the high-scale MSSM}

\author[1]{\large Gino Isidori\thanks{isidori@physik.uzh.ch}}
\author[1]{\large Sokratis Trifinopoulos\thanks{trifinos@physik.uzh.ch}}
\affil[1]{\emph{\normalsize Physik-Institut, Universit\"at Z\"urich, CH-8057 Z\"urich, Switzerland}}

\date{}

\published{
\begin{flushright}

\end{flushright}
\vskip 2cm }

\begin{document} 

\maketitle

\begin{abstract} \noindent
We analyse the sensitivity of quark flavour-changing observables to the MSSM, in a regime of heavy superpartner masses. We analyse four distinct and motivated  frameworks characterising the structure of the soft-breaking terms by means of approximate flavour symmetries. We show that a set of six low-energy observables with realistic chances of improvement in the near future, namely $\Delta M_{s,d}$,  $\epsilon_K$, $\epsilon_K'/\epsilon_K$, $\mathcal{B} (K\to \pi\nu\bar\nu)$, and the phase of $D$--$\bar D$ mixing, could play a very important role in charactering these frameworks in a regime of superpartner  masses up to $\mathcal{O}(100)$~TeV. We show that these observables remain very interesting even in a long-term perspective, i.e.~even taking into account the direct mass reach of the most ambitious future high-energy colliders.
\end{abstract}

\clearpage

\section{Introduction}

The Minimal Supersymmetric extension of the Standard Model (MSSM) is one of the most
motivated and attractive  ultraviolet completions of the Standard Model (SM).
The absence of direct signals of new particles at the LHC have pushed the scale 
of Supersymmetry (SUSY) breaking in the few TeV regime (or above), 
making the MSSM a less natural solution to the electroweak hierarchy problem.
However, many other virtues of the MSSM, such as gauge coupling unification, 
a natural dark-matter candidate, and the possibility of coherently 
embedding  gravitational interactions in the context of supergravity, remains open
if the SUSY breaking scale is up to $\mathcal{O}(100)$~TeV.
The MSSM with soft-breaking terms in this energy domain provides also 
a successful prediction for the Higgs boson mass~\cite{Giudice:2011cg,Arvanitaki:2012ps,Bagnaschi:2014rsa,Vega:2015fna,Ellis:2017erg}, as well as natural a-posteriori justification for the  heaviness of the top mass,
responsible for the radiative breaking of the electroweak symmetry~\cite{Ibanez:1982fr,Ellis:1983bp,Gamberini:1989jw,Drees:1991ab,Isidori:2017hac}. 

An additional important virtue of high-scale MSSM is a minor tension with flavour-physics observables, 
or a less severe SUSY flavour problem~(see e.g.~\cite{Gabbiani:1996hi,Misiak:1997ei,Abel:2001vy,Isidori:2010kg}).
However, a residual flavour problem remains and flavour-physics observables may represent the only option 
to test the MSSM in the near future, if the SUSY breaking scale lies is in the $10-100~ \rm TeV$ domain.
Such high-scale breaking, which is quite motivated given the argument listed above, would indeed 
prevent direct signals of super-partners at the LHC, even during the forthcoming high-luminosity phase, 
while indirect signals may still be within the reach of existing facilities. 
The purpose of this paper is to provide a quantitative estimate, and a detailed comparison, of the sensitivity 
of selected flavour-physics observables up to $\mathcal{O}(100)$~TeV.

The fact that MSSM could, in general terms, induce sizeable deviations from the SM in specific flavour-changing 
observables, such as $\epsilon_K$, even with squark masses above 100~TeV, 
is quite clear from the general sensitivity studies in the literature (see e.g.~\cite{Gabbiani:1996hi,Misiak:1997ei}).
What is less obvious are the following two questions: i) does this statement remains true under realistic 
hypothesis about the flavour structure of the soft-breaking terms, able to justify (at least to  some extent)
why we have not seen any
deviation form the SM yet? ii) Which low-energy observables --with realistic chance of improvements in the near future--
are still interesting, given the existing tight  constraints on most of them? iii)  Do flavour observables remains interesting 
even in a long-term perspective, i.e.~taking into account direct mass reach on squarks of future colliders such as the 
FCC-ee and the FCC-hh~\cite{Mangano:2017tke,Golling:2016gvc}.
These are the basic questions we aim to address in the present study.

To achieve this goal, we consider four basic hypotheses about the flavour structure of the soft-breaking terms: the Minimal Flavour Violation (MFV) hypothesis~\cite{Chivukula:1987py,Hall:1990ac,DAmbrosio:2002vsn}, a chiral $U(2)$ flavour symmetry~\cite{Barbieri:1995uv, Barbieri:1996ww,Barbieri:2011ci}, a $U(1)$ symmetry a la  Froggatt-Nielsen~\cite{Froggatt:1978nt,Lalak:2010bk}, and finally the framework of disoriented $A$ terms~\cite{Giudice:2012qq}. As we shall 
discuss, implementing these different hypotheses leads to a MSSM with a different degree of tuning in the flavour sector.
However, it is fair to state that all of them address in a satisfactory way the residual SUSY flavour problem,
i.e.~provide a reasonable a-posteriori justification of why we have not seen yet any 
non-standard effect at low energies. 

Within these four basic frameworks we analyse a set of representative $\Delta F=2$ and ${\Delta F=1}$ flavour-changing observables
in the quark  sector. 
Restricting the attention to those where we can expect significant improvements on the experimental and/or the 
theoretical side, we address the pragmatic question, if a significant ($3\sigma$) deviation from the SM --compatible with 
present constraints-- can be generated within the MSSM. We address this question as
a function of the soft-breaking scale, determining this way the maximal scale at which each of the flavoured versions of the 
MSSM listed above can be tested indirectly via low-energy flavour-changing observables.

\section{Flavour structure the soft-breaking terms}
\label{sec:flavour_models}

\subsection{Generalities}
We consider the MSSM with $R$ parity conservation, whose soft-breaking terms 
can be divided into three categories,
\begin{enumerate}[i)]
	\item mass terms for the scalar fields:
	\begin{align}
			&- m_{H_U}^2 H_i^{U*} H_i^U - m_{H_D}^2 H_i^{D*} H_i^D - (m_{\tilde{Q}}^2)_{IJ} \tilde{Q}_i^{I*}\tilde{Q}_i^{J} - (m_{\tilde{U}}^2)_{IJ} \tilde{u}_R^{J *}\tilde{u}_R^{I} - (m_{\tilde{D}}^2)_{IJ} \tilde{d}_R^{J *}\tilde{d}_R^{I}\notag \\
			&-(m_{\tilde{L}}^2)_{IJ} \tilde{L}_i^{I*}\tilde{L}_i^{J} - (m_{\tilde{E}}^2)_{IJ} \tilde{e}_R^{+ I *		}\tilde{e}_R^{+ J} 
			\label{eq:softmasses}
\end{align}
	\item mass terms for the gauginos:
	\begin{equation}
		\frac{1}{2} M_1 \lambda_B\lambda_B + \frac{1}{2} M_2 \lambda_A^i\lambda_A^i + \frac{1}{2} M_3 \lambda_G^\alpha\lambda_G^\alpha + \text{h.c.}
	\end{equation}
	\item trilinear couplings of the scalar fields:
	\begin{equation}
			\epsilon_{ij} (A_U)_{IJ} H_i^U \tilde{Q}_{j}^I  \tilde{u}_R^{J *} + \epsilon_{ij} (A_D)_{IJ} H_i^D \tilde{Q}_{j}^I \tilde{d}_R^{J *}  + \epsilon_{ij} (A_L)_{IJ} H_i^D \tilde{L}_{j}^I \tilde{e}_R^{J *} + \text{h.c.}
	\end{equation}
\end{enumerate}
Beside the two Higgs mass terms and the three gaugino masses, the vast majority of soft breaking terms 
(hence of the free parameters of the model) are encoded in the $3 \times 3$ complex matrices
$m_{\tilde{Q}}^2$, $m_{\tilde{U}}^2$, $m_{\tilde{D}}^2$, $m_{\tilde{L}}^2$, $m_{\tilde{E}}^2$, $A_U$ and $A_D$,
characterising the flavour structure of the model.

Since we assume the overall scale of the soft-breaking terms to be significantly higher that the electroweak scale, it is a good approximation
to neglect the electroweak corrections to the physical masses of squarks and leptons.  This allow us to establish a 
simple connection between soft-breaking terms (i.e.~Lagrangian parameters) and the $6\times 6$ squark and slepton 
mass matrices. 
For example, in the case of the up squarks, the $6\times 6$  mass matrix reads
\begin{equation}
	\tilde{M}_{U}^2 \approx \left( {\begin{array}{*{20}{c}}
	(\tilde{M}_{U}^2)_{LL}&(\tilde{M}_{U}^2)_{LR} \\
	(\tilde{M}_{U}^2)_{LR}&(\tilde{M}_{U}^2)_{RR}
	\end{array}} \right)~,
	\label{eq:down-squark mass matrix}
\end{equation}
where
\begin{align}
	(\tilde{M}_{U}^2)_{LL}&=m_{\tilde{Q}}^2~, \notag \\
	(\tilde{M}_{U}^2)_{LR}&= v_U A_U~, \notag \\
	(\tilde{M}_{U}^2)_{RR}&=m_{\tilde{U}}^2~.
	\label{eq:M_D_entries}
\end{align}
Similar relations holds for  the down-type ($\tilde{M}_{D}^2$) and the slepton ($\tilde{M}_L^2$) mass matrices.

A completely generic (flavour-anarchic) structure for the soft-breaking terms leads to strong tensions with various
flavour-changing neutral-current (FCNC) and of CP-violating observables, even for sfermion masses well above 10~TeV
(see e.g.~\cite{Gabbiani:1996hi,Misiak:1997ei,Abel:2001vy,Isidori:2010kg}). This is why in the following we consider 
four different hypothesis about the underlying flavour structure of the theory able to provide a consistent (a posteriori)
justification of why the flavour off-diagonal 
entries in (\ref{eq:down-squark mass matrix}), and the other sfermion mass matrices, are suppressed.

\subsection{Minimal Flavour Violation}

The starting point of the MFV hypothesis is flavour group $\mathcal{G}_{F}$,
\begin{align}
	\mathcal{G}_{F} = \mathcal{G}_{q} &\times \mathcal{G}_{l} \notag \\
	\mathcal{G}_{q} = U(3)_{Q} \times U(3)_{U} \times U(2)_{D}, \ \ \ & \ \ \ \mathcal{G}_{l} = U(3)_{L} \times U(3)_{E}~,
	\label{eq:SM flavour symmetry} 
\end{align}
corresponding to the flavour symmetry that the SM Langrangian enjoys in the limit of vanishing Yukawa couplings. 
The basic assumption is that the only quantities that break this symmetry are spurion fields proportional to the SM Yukawa couplings themselves. In the MSSM context this implies that the soft-breaking terms can be reconstructed entirely out of 
appropriate powers of the SM Yukawa matrices $Y_{U,D,E}$~\cite{DAmbrosio:2002vsn}\footnote{Without loss of generality, 
we work in the so-called down-basis, where the SM Yukawa couplings assume the following from:
$Y_U = \lambda_U V$, $Y_D = \lambda_D$, and $Y_E = \lambda_E$. Here 
$\lambda_U=\text{diag}(y_u,y_c,y_t)$, $\lambda_D=\text{diag}(y_d,y_s,y_b)$, $\lambda_E=\text{diag}(y_e,y_{\mu},y_{\tau})$ and $V$ is the CKM matrix.}.
Keeping only the leading terms in the expansion in terms of flavour-changing spurions leads to a significant reduction in the number of free
parameters~\cite{Colangelo:2008qp, AbdusSalam:2014uea}. Furthermore, without affecting the phenomenological analysis
(focused on quark flavour-changing observables), 
we assume real gaugino masses and universal slepton masses. According to these hypotheses, we end up with a 
total of 14  parameters characterising all the  soft-breaking in this framework:
\begin{align}
M_1, \ \ &M_2, \ \ M_3  \notag \\
(\tilde{M}_{Q}^2)_{IJ} = \tilde{m}_Q^2 \left( \delta_{IJ} + x_1 V_{3I}^*V_{3J} \right), \ \ \ (\tilde{M}_{U}^2)_{IJ} &= \tilde{m}_U^2 \left( \delta_{IJ} + x_2 \delta_{3I} \delta_{3J} \right), \ \ \ (\tilde{M}_{D}^2)_{IJ} = \tilde{m}_D^2 \delta_{IJ} \notag \\
(\tilde{M}_{L}^2)_{IJ} = \tilde{m}_L^2 \delta_{IJ},& \ \ \ (\tilde{M}_{E}^2)_{IJ} = \tilde{m}_E^2 \delta_{IJ} \notag \\
(A_U)_{IJ} = \tilde{a}_0 \delta_{3I}V_{3J}, \ \ \ (A_D)_{IJ} = \tilde{a}_0 &\left(\delta_{3I} \delta_{3J} + y_5 \delta_{3I} V_{3J} \right), \ \ \ (A_E)_{IJ} = \tilde{a}_0 \delta_{IJ}~.
\label{eq:MSSM-11} 
\end{align}
The adimensional $x_i$ and the overall soft masses $\tilde{m}_{Q/U/D}^2$ and $\tilde{m}_{L/E}^2$ are necessarily real, 
while the remaining 
parameters can be complex. By consistency of the framework,  the dimensionless parameters $x_i$ and $y_i$ 
are restricted to be at most of $\mathcal{O}(1)$~\cite{DAmbrosio:2002vsn}.

\subsection{$U(2)$ chiral flavour symmetry}
An approach quite similar to the MFV hypotheses, but based on a smaller flavour symmetry and a larger set of symmetry-breaking terms, is obtained assuming a flavour symmetry acting only on the light generations. The basic premise of this hypothesis is the 
observation that $U(2)$ flavour symmetries provide a good description of the SM Yukawa couplings, explaining the smallness 
of light-fermion masses compared to third-generation ones. Restricting the attention to the quark sector, we employ the 
following flavour symmetry~\cite{Barbieri:2011ci}
\begin{equation}
	\mathcal{G}_{q} = U(2)_Q \times U(2)_U \times U(2)_D~,
\end{equation}
acting on the first two generations. We further assume that the symmetry is broken by three spurions:
the leading spurion $V_Q$, responsible for breaking the left-handed flavour subgroup $U(2)_Q$ (or the  left-handed  mixing between
third- and light-generations)\footnote{The natural size of the leading spurion is $V_Q = \mathcal{O}(|V_{ts}|)$.},  and the sub-leading 
spurions $\Delta Y_{U(D)}$,  transforming as bi-doublets 
of $U(2)_{U(D)} \times U(2)_Q$.
 In terms of these symmetry breaking terms, the quark Yukawa couplings can be written as
\begin{equation}
	Y_U = y_t \left(
			\begin{array}{c;{2pt/2pt}c}
					\Delta Y_U &  x_t V_Q \\ \hdashline[2pt/2pt]
					0 & 1 
			\end{array}
	\right)  \ \ \ \text{and} \ \ \ 
	Y_D = y_b \left(
			\begin{array}{c;{2pt/2pt}c}
					\Delta Y_D & x_b V_Q \\ \hdashline[2pt/2pt]
					0 & 1
			\end{array}
	\right)~,
\end{equation}
where $y_{t,b}$ are the third-generation Yukawa couplings, and $x_{t,b}$ are $\mathcal{O}(1)$ parameters.

When describing flavour-mixing in the soft-breaking terms, we neglect the subleading $\Delta Y_{U(D)}$ spurions,
which are very small as a consequence of  the smallness of light-quark masses.
In this limit flavour mixing appears only in the left-handed sector.  
Employing the explicit parametrization given in~\cite{Barbieri:2011ci}, one may express the squark mass matrix in terms of a small number of mixing angles and masses:
\begin{align}
	\tilde{M}_{Q}^2 = W_L^d  \times \text{diag} (\tilde{m}_{Q_h}^2&, \tilde{m}_{Q_h}^2, \tilde{m}_{Q_l}^2) \times W_L^{d \dagger},  \notag \\
\tilde{M}_{U}^2 = \text{diag} (\tilde{m}_{u_h}^2, \tilde{m}_{u_h}^2, \tilde{m}_{u_l}^2), \ \ & \	\tilde{M}_{D}^2 = \text{diag} (\tilde{m}_{d_h}^2, \tilde{m}_{d_h}^2, \tilde{m}_{d_l}^2), \notag \\
	A_U = a_0, \ \ & \	A_D = a_0 y_b~. \notag \\		
	\label{eq:soft-breaking terms U(2)} 
\end{align}
where
\begin{equation}
W_L^d = \left( {\begin{array}{*{20}{c}}
	c_d&\kappa^*&-\kappa^* s_L e^{i\gamma}\\
	-\kappa&c_d&-c_d s_L e^{i\gamma}\\
	0&s_L e^{-i\gamma}&1
	\end{array}} \right)~.
\end{equation}
The parametric size of the parameters appearing in the left-handed mixing matrix  $W_L^d$ is as following:
\begin{equation}
c_d \sim \mathcal{O}(1)~, \qquad  \kappa=c_d \frac{V_{td}}{V_{ts}} \sim \lambda \equiv |V_{ts}|~, 
\qquad 
 \left|s_L e^{\pm i\gamma}\right| \sim \lambda^2~.
 \end{equation} 
To further simplify this framework, and also to distinguish it from the general MFV case discussed above, we further assume 
the limit where the first two generations of squarks are very heavy.
This hypothesis is not an exclusive requirement of the $U(2)$ flavour symmetry, but it is a particularly interesting case considered in the literature (the so-called effective SUSY or split-family SUSY~\cite{Dimopoulos:1995mi,Cohen:1996vb}),
which cannot be achieved under the MFV hypothesis and 
which is particularly motivated given the present lack of direct SUSY signals at the LHC 
(see e.g.~\cite{Buckley:2016kvr} and references therein). 
Under this additional hypothesis our $U(2)$ framework is fully characterised by the entries in $W_L^d$, the universal
trilinear term $a_0$, and by the light third-generation masses  ($\tilde{m}_{Q_l}^2$, $\tilde{m}_{u_l}^2$,  $\tilde{m}_{d_l}^2$,
and corresponding terms for the sleptons).

\subsection{$U(1)$ Froggatt-Nielsen}
\label{sec:U(1)}
As representative example of a  framework with larger flavour-violating terms, we consider
the holomorphic $U(1)$ Froggatt-Nielsen symmetry acting on the quark sector proposed in~\cite{Lalak:2010bk}.
  In order to successfully reproduce the observed SM mass hierarchies, the quarks (and correspondingly to the squarks) 
  are assigned the following $U(1)_{\rm FN}$ charges:
\begin{equation}
  Q_{L\text{ }1,2,3} \sim (3,2,0)~, \ \ \ u_{R\text{ }1,2,3}^c \sim (3,2,0)~, \ \ \ d_{R\text{ }1,2,3}^c \sim (1,0,0)~.
	\label{eq:U(1) charges}
\end{equation}
The symmetry is spontaneously broken via one single familon $\theta$ carrying a positive $U(1)_{\rm FN}$ charge $Q_\theta=+1$. 
For instance, the up-Yukawa coupling takes the form
\begin{equation}
\epsilon_{ij} Y_U^{IJ} H_i^U Q_{j}^I  (u^c)^J = \epsilon_{ij} \left[ a_{IJ} \left( \frac{\theta}{M}\right)^{u_J+q_I}\right] H_i^U Q_{j}^I  (u^c)^J~,
\end{equation}
where $a_{IJ}$ are coefficients of $\mathcal{O}(1)$, $\theta$ denotes the familon VEV, $q_I$, $u_I$ are the charges defined in (\ref{eq:U(1) charges}) and $M$ the cut-off scale of the effective theory. 
Setting $\epsilon = \theta / M \approx \lambda$ one obtains a good description of the SM Yukawa coupling 
(with suitable choices of the  $\mathcal{O}(1)$ parameters).

Proceeding in a similar manner for the soft-breaking terms in the squark sector 
we arrive to the following parametric decomposition  (we omit to write the unknown $\mathcal{O}(1)$ complex coefficients):
\begin{align}
\label{eq:U1_mass_terms}
&\tilde{M}_{Q}^2 = \tilde{m}_Q^2 \left( {\begin{array}{*{20}{c}}
	1&\epsilon&\epsilon^3\\
	\epsilon&1&\epsilon^2\\
	\epsilon^3&\epsilon^2&1
	\end{array}} \right), \ \ \
\tilde{M}_{U}^2 = \tilde{m}_U^2 \left( {\begin{array}{*{20}{c}}
	1&\epsilon&\epsilon^3\\
	\epsilon&1&\epsilon^2\\
	\epsilon^3&\epsilon^2&1
	\end{array}} \right), \ \ \
\tilde{M}_{D}^2 = \tilde{m}_D^2 \left( {\begin{array}{*{20}{c}}
	1&\epsilon&\epsilon\\
	\epsilon&1&1\\
	\epsilon&1&1
	\end{array}} \right), \notag \\
&A_U = a_0 \left( {\begin{array}{*{20}{c}}
	\epsilon^6&\epsilon^5&\epsilon^3\\
	\epsilon^5&\epsilon^4&\epsilon^2\\
	\epsilon^3&\epsilon^2&1
	\end{array}} \right), \ \ \
	A_D = a_0 \left( {\begin{array}{*{20}{c}}
	\epsilon^4&\epsilon^3&\epsilon^3\\
	\epsilon^3&\epsilon^2&\epsilon^2\\
	\epsilon&1&1
	\end{array}} \right)~.
\end{align}
The $\mathcal{O}(1)$ complex coefficients of the $\mathcal{O}(\epsilon^n)$ terms in these matrices, together with their overall normalisation, represent the free parameters of this framework. As in the MFV case we assume universal slepton masses.

\subsection{Disoriented $A$ terms}

The disoriented $A$-term framework~\cite{Giudice:2012qq} is a scenario where flavour violation occurs 
only in the $L$--$R$ mixing terms, hence the SM Yukawa and the trilinear soft-breaking terms which,
however, are not assumed to be aligned, as opposed to the MFV case. More precisely, the soft-breaking mass terms 
in (\ref{eq:softmasses}) are assumed to be exactly diagonal, whereas the $A$ terms can be 
decomposed as
\begin{align}
&(A_{F})_{IJ} = A_0 \theta_{IJ}^{F} y_{F_J}~, \ \ \ \ F=U,D \notag \\
&(A_{E})_{IJ} = A_0 \theta_{IJ}^{E} y_{E_J}~.
\end{align}
where $\theta_{IJ}^{F/E}$ are generic $\mathcal{O}(1)$ mixing angles. The latter, together with the universal soft masses, 
are the free parameters of this framework.

The possibility of the absence of flavour violation in the soft masses (quadratic terms), together with a sizeable 
flavour mixing in the $A$ terms is not unlikely in supersymmetric theories: a scenario of this type
 can be realised, for instance, if a non-abelian flavour symmetry act on the $R$-invariant part of the supersymmetry-breaking terms, ensuring (total or partial) universality of soft masses, but  is violated in the $R$-charged sector, allowing for general trilinear terms. It is 
also worth to stress that the separation between soft masses and trilinear interactions of the first two generations is 
quite robust under the renormalization-group flow, making this scenario technically stable. Last but not least, the 
experimental bounds on the $\theta_{IJ}^{F/E}$ are indeed of $\mathcal{O}(1)$, even for squark masses of a few TeV~\cite{Giudice:2012qq} making this scenario quite consistent from the phenomenological point of view.

\section{Flavour-changing observables} 
\label{sec:observables}

In this Section we enumerate  the flavour-changing observables taken into account in our phenomenological analysis, 
illustrating their key features both in the SM and in the MSSM.
On general grounds, the number of potentially interesting flavour-changing observables is quite large. However, we 
restrict the attention to a relatively small number according to the following main criterium: we consider observables 
which provide, at present, a stringent constraint on the MSSM (in flavour space) and/or can be expected to 
be significantly improved in the near future. Moreover, we focus only
on lepton-flavour conserving observables.

According to this criterium, we restrict the attention 
to all the accessible $\Delta F=2$ observables and to 
three selected $\Delta F=1$ transitions, namely $\epsilon_K'/\epsilon_K$, $\mathcal{B} (B\to X_s\gamma)$
and $\mathcal{B} (K^+ \to \pi^+ \bar{\nu} \nu)$.  The latter are representative of the three most-relevant 
classes of  relevant flavour-chaning amplitudes: non-leptonic ones ($\epsilon_K'/\epsilon_K$),
FCNC with leptons ($K^+ \to \pi^+ \bar{\nu} \nu$) and FCNC with photons ($B\to X_s\gamma$).\footnote{We 
do not include in our analysis observables related to the recent $B$-physics anomalies 
(see e.g.~\cite{Ciezarek:2017yzh,Blake:2016olu}) 
since, as we have explicitly verified, there is no way to generate sizeable modifications 
to these observables the $R$-parity conserving high-scale MSSM . As shown 
in~\cite{Altmannshofer:2017poe, Trifinopoulos:2018rna,Trifinopoulos:2019lyo},
a good description of current anomalous results in $B$-physics can be obtained 
in supersymmetric extensions of the SM relaxing the hypothesis of $R$-parity conservation.}

The current values of these observables are shown in Table~\ref{tbl:flavour_obs}. In the last column of this
table we show a hypothetical future scenario where, thanks to (realistic) improvements on the experimental
and/or the theoretical side, some these observables could exhibit a significant deviation from the SM. 
These hypothetical future results will be used in our numerical analysis to assess 
the sensitivity of each observable to the high-scale MSSM.  

A more detailed discussion of the various observables follows below.

\begin{table}[t]
\addtolength{\arraycolsep}{3pt}
\renewcommand{\arraystretch}{1.2}
\centering
\begin{tabular}{|c||c|c||c|}
\hline
observable & experiment~\cite{Patrignani:2016xqp} &  $O_{\rm exp}/O_{\rm SM} - 1 $  & {\em future scenario} $(3\sigma)$ \\
\hline\hline
$\Delta M_{B_d}$ & $(0.5064 \pm 0.0019)~{\rm ps}^{-1}$ 
  &  $-0.13 \pm 0.09$~\cite{DiLuzio:2019jyq,Aoki:2019cca}   &  $-0.13 \pm 0.04$ \\
\hline
$\Delta M_{B_s}$ & $(17.757 \pm 0.021)~{\rm ps}^{-1}$ &   $-0.12 \pm 0.07$~\cite{DiLuzio:2019jyq,Aoki:2019cca} 
& $-0.12 \pm 0.04$ \\
\hline
$|\epsilon_K|$ & $(2.229\pm 0.010) \times 10^{-3}$ &  $0.10 \pm 0.09$~\cite{Alpigiani:2017lpj} 
& $\phantom{+}0.10 \pm 0.03$ \\
\hline
$\mathcal{B}(B\to X_s\gamma)$ & $(3.52\pm 0.25)\times 10^{-4}$    & $0.11  \pm  0.11$~\cite{Misiak:2006zs} &  -- \\ 
\hline
\hline
 \multicolumn{2}{|c|}{} &  $O_{\rm exp}- O_{\rm SM}$ &  \\
\hline
 \hline
$\epsilon_K'/\epsilon_K$ & $(16.6 \pm 2.3) \times 10^{-4}$  
 & $ (11  \pm  7)\times 10^{-4} $  &   $(11  \pm  3.6)\times 10^{-4} $   \\
\hline
$\Im(M^D_{12})/M_D^2$ & $(0.0 \pm 4.6)\times 10^{-17}$~\cite{Alpigiani:2017lpj}   & $(0.0 \pm 4.6)\times 10^{-17}$
 &  $(4.6 \pm 1.5)\times 10^{-17}$ 
 \\
\hline
$\mathcal{B}(K^+\to \pi^+\nu\bar\nu)$ & $(0.85 \pm 0.5)\times 10^{-10}$~\cite{Artamonov:2008qb,NA62-preliminary}    & $(0.0 \pm 0.5)\times 10^{-10}$  &  
$(0.3 \pm 0.1)\times 10^{-10}$     \\ 
\hline
$\Delta M_K/M_K$ & $ 7.0 \times 10^{-16}$ 
 & $(0 \pm 7)  \times 10^{-16}$      & --  \\
 \hline
\end{tabular}
\caption{Experimental values and SM predictions for the observables used in the numerical 
analysis (see main text).}
\label{tbl:flavour_obs}
\end{table}

\begin{itemize}
	\item $\Delta F = 2.$ If quark- and lepton-flavour violating interactions occur independently and 
	without a special chiral structure, $\Delta F = 2$ amplitudes naturally provide the most 
	stringent constraints on heavy scale new-physics (see e.g.~\cite{Fleischer:2003xx,Lenz:2010gu}).
	In the $B_d$ and $B_s$ systems both magnitudes and phases of the  
	meson-antimeson mixing amplitudes are dominated by short-distance dynamics and 
	are experimentally determined with high accuracy.  In order to minimise the impact of the 
	current SM uncertainties, we implement the corresponding constraints 
	normalising the experimental results to SM predictions. 	
	On the phases, we can expect only marginal improvements 
	in the data/theory comparison before being saturated by irreducible theory errors.  On the other hand, significant 
	room for improvement can be expected on the magnitudes, thanks to improved Lattice-QCD predictions 
	for $\Delta M^{\rm SM}_{d(s)}$ (see e.g.~\cite{DiLuzio:2019jyq,Aoki:2019cca}).		
	The situation is opposite in the Kaon and $D$-meson systems, where only the phases (or better the CP-violating 
	mixing amplitudes) are short-distance dominated. Despite the lack of a precise SM prediction for $\Delta M_K$, 
	 we implement a constraint based on $\Delta M_K$ requiring 
	non-standard contributions to this observable not to exceed, in magnitude, the SM short-distance 
	contribution estimated in~\cite{Brod:2011ty}. As far as CP violation in $D$-meson mixing 
	is concerned, we implement the constrain on $\Im(M^D_{12})$ from~\cite{Alpigiani:2017lpj}
	without normalising it to the SM expectation (assuming the latter to be negligible). 

	The MSSM contributions to the $\Delta F = 2$ observables are induced by squarks--gauginos box diagrams. 
	The dominant contribution usually comes from the gluino-squark box, whose expression in the case of degenerate squarks 
	can be found in~\cite{Gabbiani:1996hi}.
	Depending on the spectrum of the gaugino masses, chargino-squark boxes can also become sizable. The latter can be 
	found in~\cite{Crivellin:2010ys}.
	 
	\item $\epsilon_K'/\epsilon_K.$
	The strong suppression of $\epsilon_K'/\epsilon_K$ within the SM, which is not only due to the CKM hierarchy but also to 
	an accidental SM low-energy property (the so-called $\Delta I=1/2$ rule), makes this observable a particularly sensitive probe 
	of non-standard sources of both CP and flavour symmetry breaking. The main difficulty here is the uncertainty on the SM prediction.
	The value we report in the table is an educated guess taking into account the recent results 
	in~\cite{Cirigliano:2019obu,Aebischer:2019mtr,Gisbert:2018tuf,Buras:2015yba}. For a conservative implementation 
	of the present constraint on new physics, we adopt an inflated error covering all available predictions within the $1\sigma$ interval.
	In the future scenario we illustrate how the situation could change if new lattice-QCD data on the hadronic matrix element,
	which so far are still affected by large errors~\cite{Bai:2015nea},
	 would clearly establish a deviation from the SM.

Within the MSSM there are four distinct amplitudes able to generate sizable contributions to $\epsilon_K'/\epsilon_K$: 
isospin-breaking gluino-squark box diagrams~\cite{Kitahara:2016otd} (also known as Trojan penguins~\cite{Grossman:1999av}), chargino-squark box diagrams~\cite{Khalil:2001wr} and chargino-mediated \cite{Colangelo:1998pm, Endo:2016aws} as well as 
gluino-mediated \cite{Endo:2017ums} $Z$-penguin diagrams (relevant to scenarios with sizeable trilinear terms).

	\item $\mathcal{B} (B\to X_s\gamma)$. Historically  $\mathcal{B} (B\to X_s\gamma)$ has been one of the most significant 
	constraint on the structure MSSM~\cite{Barbieri:1993av}, as well as many other new physics models. We base our constraint 
	on the precise SM result in~\cite{Misiak:2006zs} which has an error comparable to the experimental one. Given irreducible 
	uncertainties associated to non-perturbative effects, this error will hardly improve in the next few years. 
	
	The MSSM constraint is obtained expanding the SM  NNLO expression for the $B\to X_s\gamma$ rate in terms of the 
	Wilson coefficients of the dipole ($C_7$) and chromo-magnetic dipole ($C_8$) operators around their SM central values, 
	as in \cite{Hurth:2008jc}. SUSY corrections are then obtained computing 
	$\Delta C_{7(8)} = C^{\rm MSSM}_{7(8)} -C^{\rm SM}_{7(8)}$ using the one-loop  
	chargino photon- and gluon-penguin amplitudes in~\cite{Barbieri:1993av}.	 
	 
	\item $\mathcal{B} (K^+ \to \pi^+ \bar{\nu} \nu)$.  This rare process may provide one of the most clean and stringent 
	constraint on flavour-changing $Z$-penguin amplitudes. Contrary to $\mathcal{B} (B\to X_s\gamma)$, here the dominant 
	error is the experimental one, which is expected to significantly improve in the near future. 
	We base our present constraint on  
         the old BNL-E949 experimental result~\cite{Artamonov:2008qb},
	and the very recent NA62 result presented at KAON 2019 \cite{NA62-preliminary}. Averaging the two we obtain 
	the value reported in Table~\ref{tbl:flavour_obs}, whose central value coincides with the most up-to-date SM
	prediction~\cite{Buras:2015yca}.

	The relevant MSSM contributions include chargino and gluino $Z$-penguin diagrams with non-negligible, $SU(2)_L$-breaking trilinear couplings as well as chargino boxes. The former yielding the dominant contribution at high energies~\cite{Tanimoto:2015ota}.
	
\end{itemize}

\section{Numerical analysis and discussion}
\label{sec:analysis}

\subsection{Analysis strategy}
As illustrated in Table~\ref{tbl:flavour_obs},
the observables are divided into two categories depending if they are normalised to the SM or not. 
In these two groups we define the expected 
deviations from the SM as
\begin{equation}
\Delta O_i^{\rm NP}(x) = \left\{ \begin{array}{l}  O_i^{\rm NP}/O_i^{\rm SM} - 1  \\   O_i^{\rm NP}-  O_i^{\rm SM}  \end{array} \right.
\end{equation} 
where  $O_i^{\rm NP}$ denotes the observables in the MSSM, and $x$ generically denotes the corresponding 
parameters (soft masses) in a given MSSM implementation. To compute explicitly the $\Delta O_i^{\rm NP}(x)$ 
we use the expressions of the observables 
in terms Wilson Coefficients of  dimension-six operators according to the following references:
\begin{itemize}
	\item $\Delta B = 2$ and $\Delta S=2$ amplitudes:  Eqs.~(7.24)--(7.27) in Ref.~\cite{Buras:2001ra}; 
	\item $\Delta C = 2$ amplitude: Eq.~(3) in Ref.~\cite{Carrasco:2014uya};
	\item $\epsilon_K'/\epsilon_K$: Eq.~(82) in Ref.~\cite{Kitahara:2016nld},
	\item $\mathcal{B} (B\to X_s\gamma)$: Eq. (33) in Ref.~\cite{Hurth:2008jc};
	\item	$\mathcal{B} (K^+ \to \pi^+ \bar{\nu} \nu)$: Eq. (24) in Ref.~\cite{Tanimoto:2015ota}\footnote{We have contacted the authors and confirmed that the parameter $\kappa$ should be rescaled by a factor of $\left(\frac{2^{3/2} \pi \sin^2\theta_W v}{\alpha}\right)^2$.}.
\end{itemize}
The  expression of the Wilson Coefficients in the MSSM are computed at the one-loop level, taking into account all the amplitude discussed in Section~\ref{sec:observables}. The corresponding loop functions are expanded around the diagonal entries of the squark mass matrices
(mass insertion approximation) up to second non-trivial order  for each observable (see Appendix in~\cite{Colangelo:1998pm}). 
The entries in the diagonal are kept non-degenerate.  We have explicitly cross-checked our analytic one-loop expressions with the corresponding degenerate limits reported in the literature (see references listed in Section~\ref{sec:observables}). 

\medskip

In the third column of Table~\ref{tbl:flavour_obs} we report  the present constraints on the $\Delta O_i^{\rm NP}(x)$, taking into account 
current experimental results and SM estimates. We assume a Gaussian distribution with mean value
$\mu_i$ and standard deviation $\sigma_i$ (including both experimental and theoretical errors). 
In the fourth column we report hypothetical $\{\mu^F_i, \sigma^F_i\}$ corresponding to a possible 
$3 \sigma$ deviation from the SM compatible with present data.  Since no significant improvement is expected in the 
near future on the SM predictions for $\Delta M_K$ and $\mathcal{B} (B\to X_s\gamma)$, no future scenario is considered for these two observables.

Our goal is to scrutinize the capability of each flavour model of Section \ref{sec:flavour_models} to provide a best-fit point that improves over the SM setting one of the observables to the future scenario, while keeping all the others to the current values.  
For each scenario we thus construct six different $\chi^2$ functions,
\begin{equation}
\chi_i^2(x) = \left(\frac{\Delta O_i^{\rm NP}(x) - \mu^F_i}{\sigma^F_i}\right)^2+ \sum_{j\not=i} \left(\frac{\Delta O_j^{\rm NP}(x) - \mu_j}{\sigma_j}\right)^2~; \quad \quad i=1,...,6~,
\end{equation}
depending on which observable is set to  the value in the fourth column.
The $\chi_i^2(x)$ are then minimized as a function of the model parameters. 
In each case we minimize the $\chi^2$ function setting all SUSY masses around an overall scale $M$, requiring the trilinear couplings  to be compatible with the vacuum stability bounds~\cite{Casas:1996de}. As a convention, we choose $M$ to be the mass of the third generation squark and we allow the other masses to vary within one order of magnitude around this scale (except for the heavy squark masses in the $U(2)$ case, which are always assumed to be decoupled),
 checking the compatibility with bounds from direct searches. We start from the current lower bounds for the masses in~\cite{Patrignani:2016xqp} and repeat the procedure for increasing values of $M$ until we eventually reach the point of decoupling.
For each observable we interpolate between the consecutive best-fit points of each flavour model and different values of $M$ 
obtaining the continuum lines shown in Figure \ref{fig:plots}.

As far as the dimensionless parameters of the models are concerned, they are  varied within their natural range as indicated in Section \ref{sec:flavour_models}. 
More precisely, the absolute size of the $\mathcal{O}(1)$ parameters is allowed to vary between $1/10$ and $4$ in the MFV and $U(2)$ cases, 
and between $1/10$ and $1$ in the other two. 
A special care is required for the  $U(1)_{\rm FN}$ model, in the regime $M <10$~TeV: here the 
 $\mathcal{O}(1)$ parameters controlling first-second generation mass insertions  need to be tuned at the per-cent level  (or even per-mil level, 
for the lowest values of $M$) in order to satisfy the relevant low-energy bounds.
In order to present results for this model at low $M$ we allow this tuning. 
Moreover,  to clearly distinguish the $U(1)_{\rm FN}$ model from the disoriented $A$-terms scenario, 
we set off-diagonal trilinear couplings to zero in the former case.

\subsection{Discussion}

The lines interpolating the consecutive best-fit points for the six flavour observables as a function of $M$, in the different flavour models,
are shown in Figure \ref{fig:plots}. The shapes of the lines suffer from some minor numerical instabilities. Still, they provide a good illustration of the main findings of this analysis, which can be summarised as follows:

\begin{figure}[p]
\centering
\includegraphics[width=8.44cm]{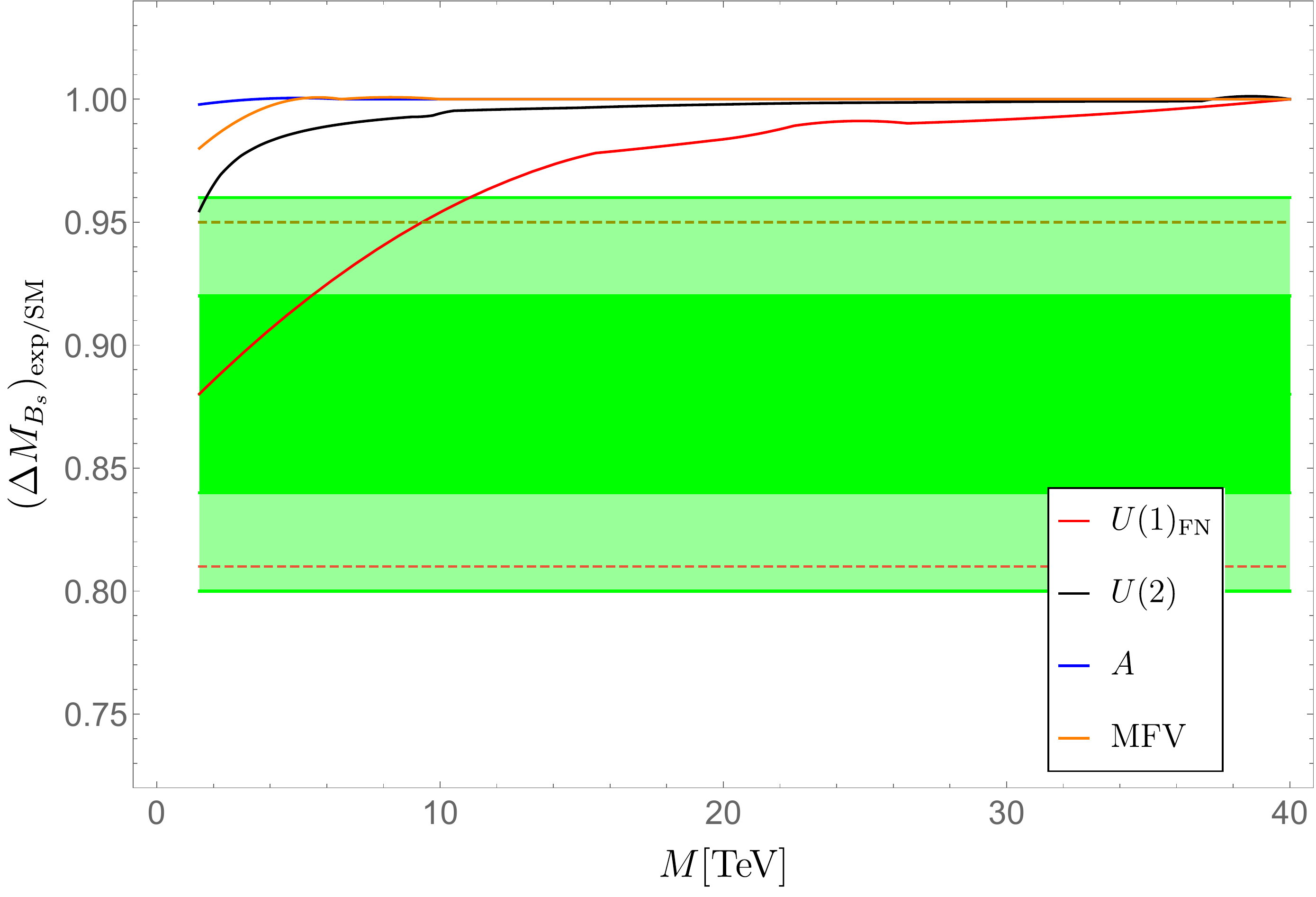} 
\includegraphics[width=8.44cm]{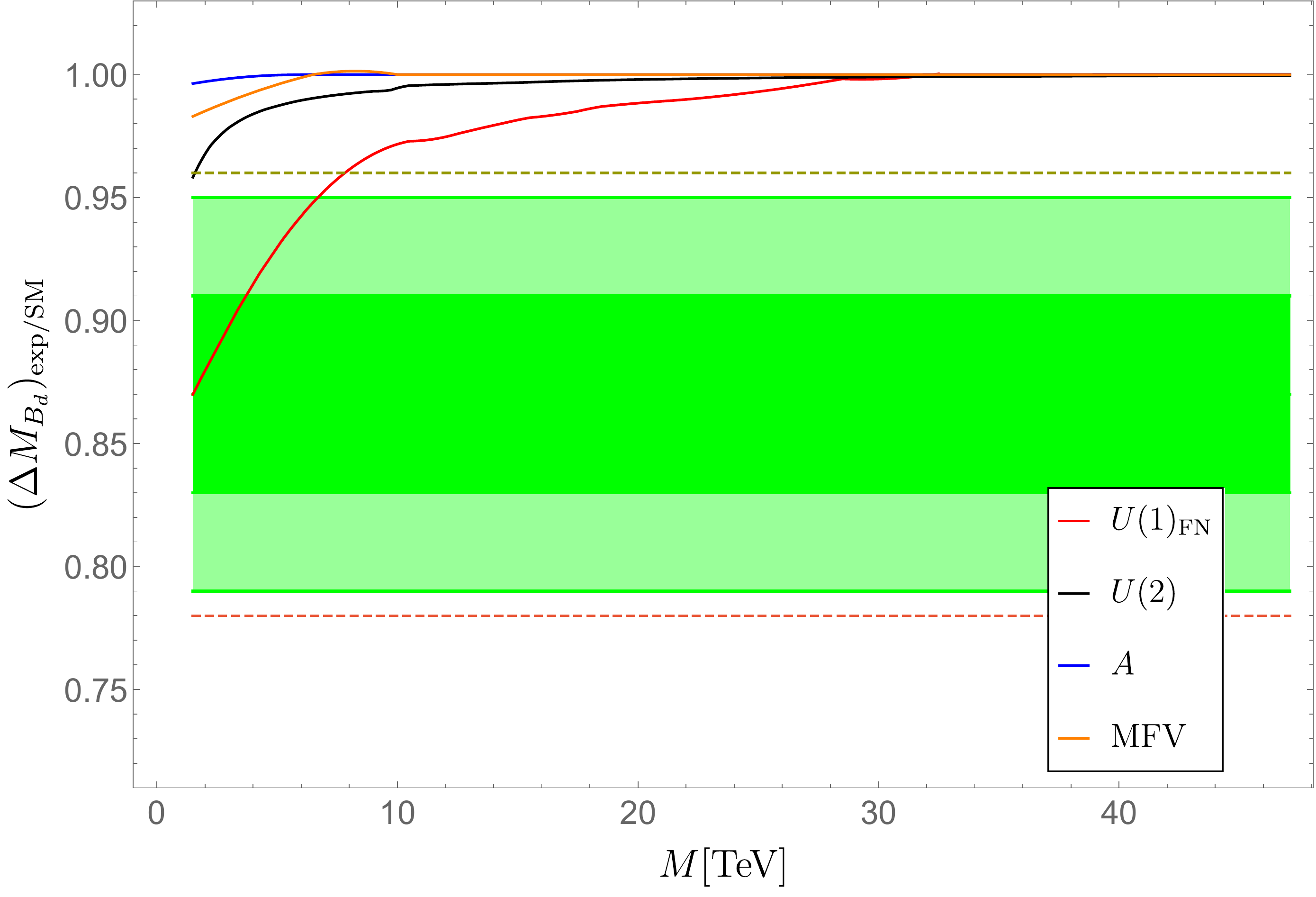} \\[5pt]
\includegraphics[width=8.44cm]{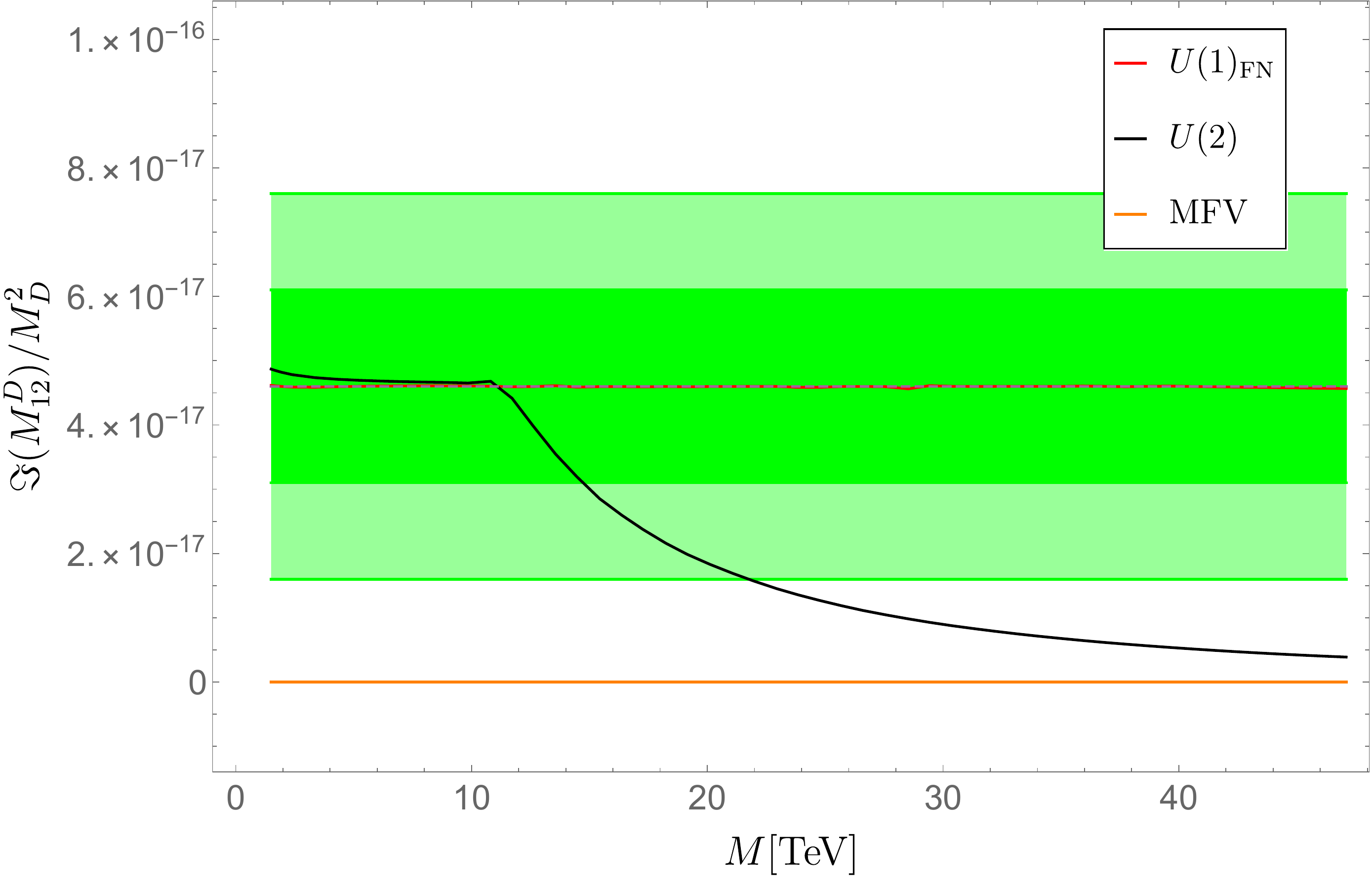} 
\includegraphics[width=8.44cm]{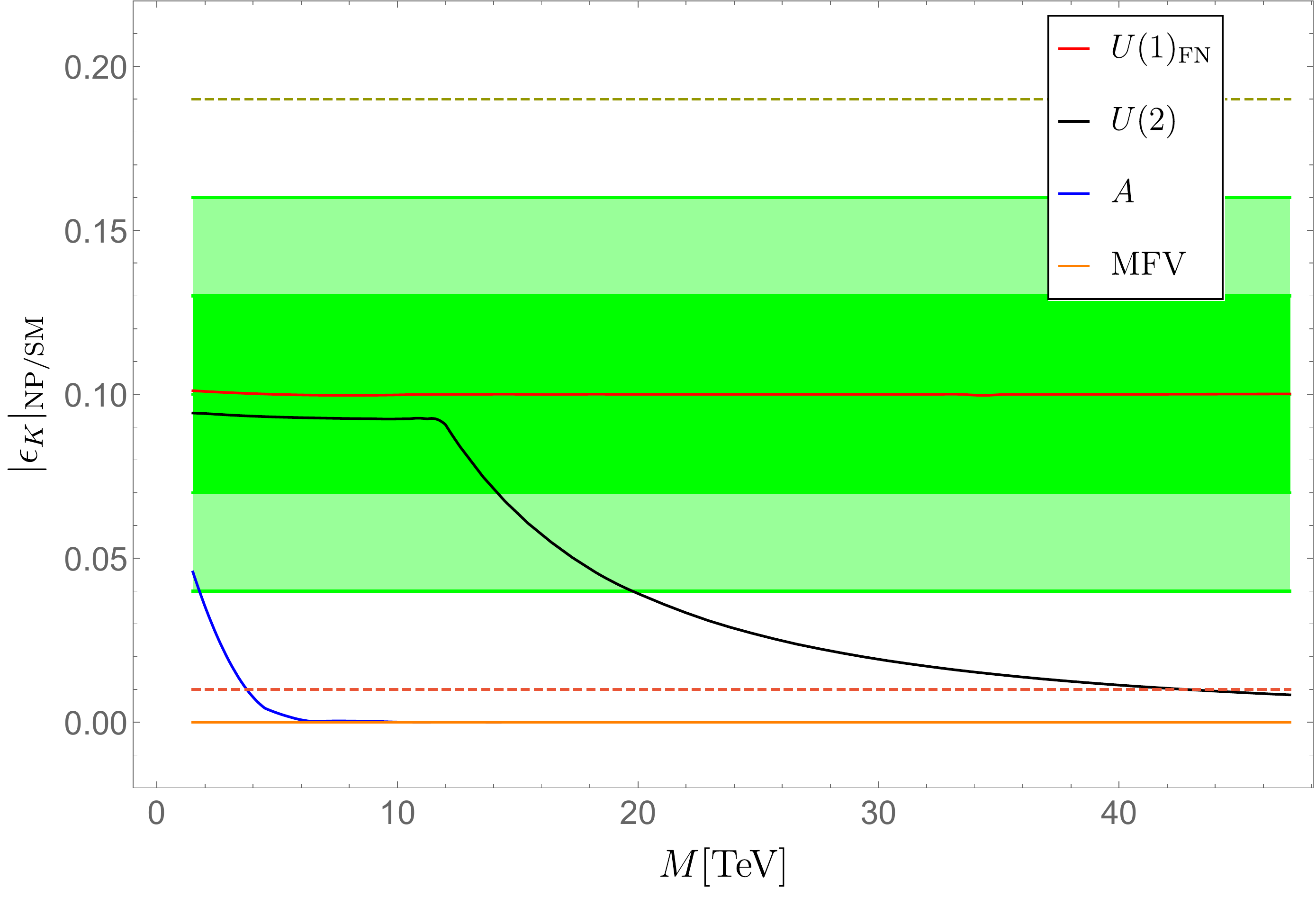} \\[5pt]
\includegraphics[width=8.44cm]{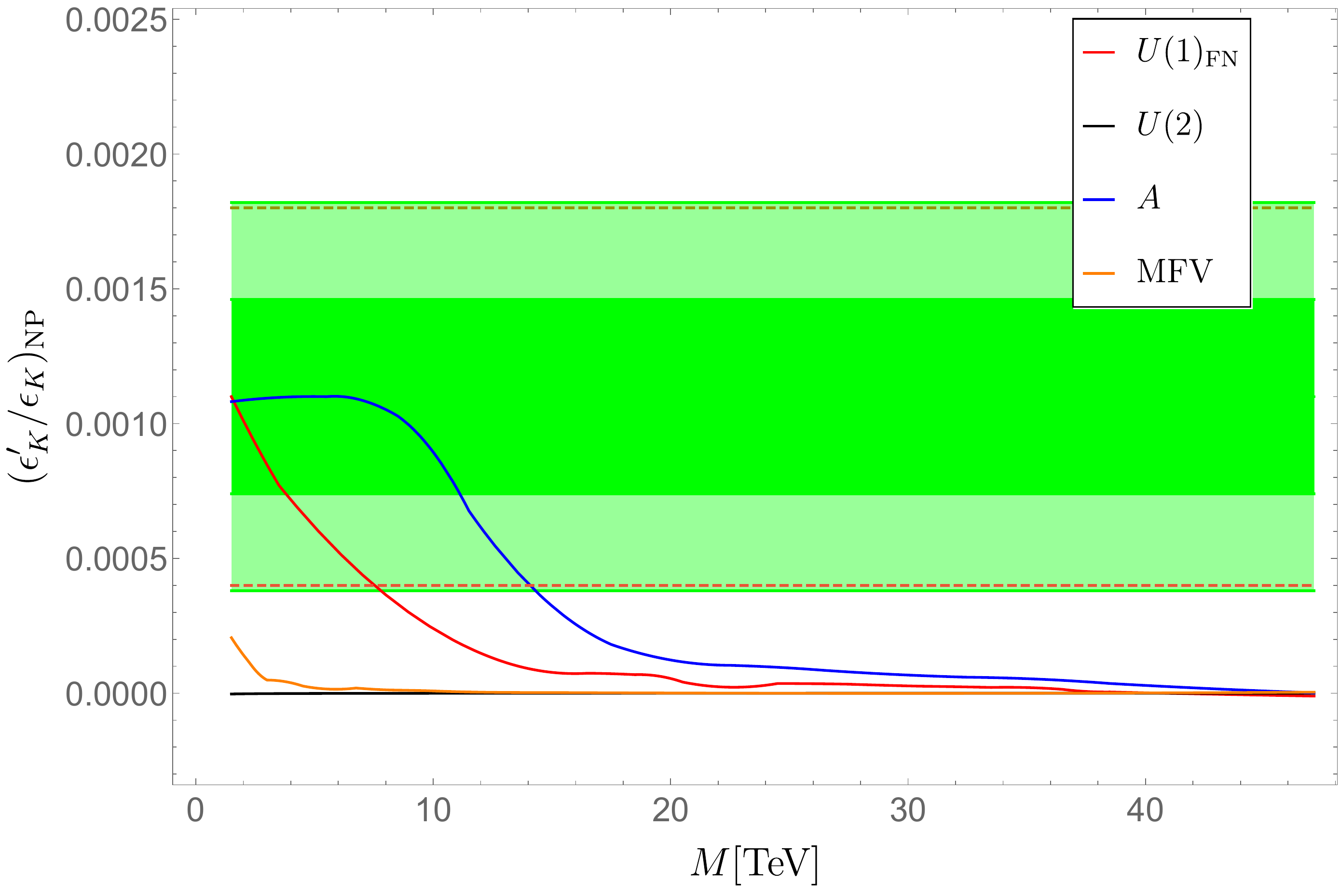} 
\includegraphics[width=8.44cm]{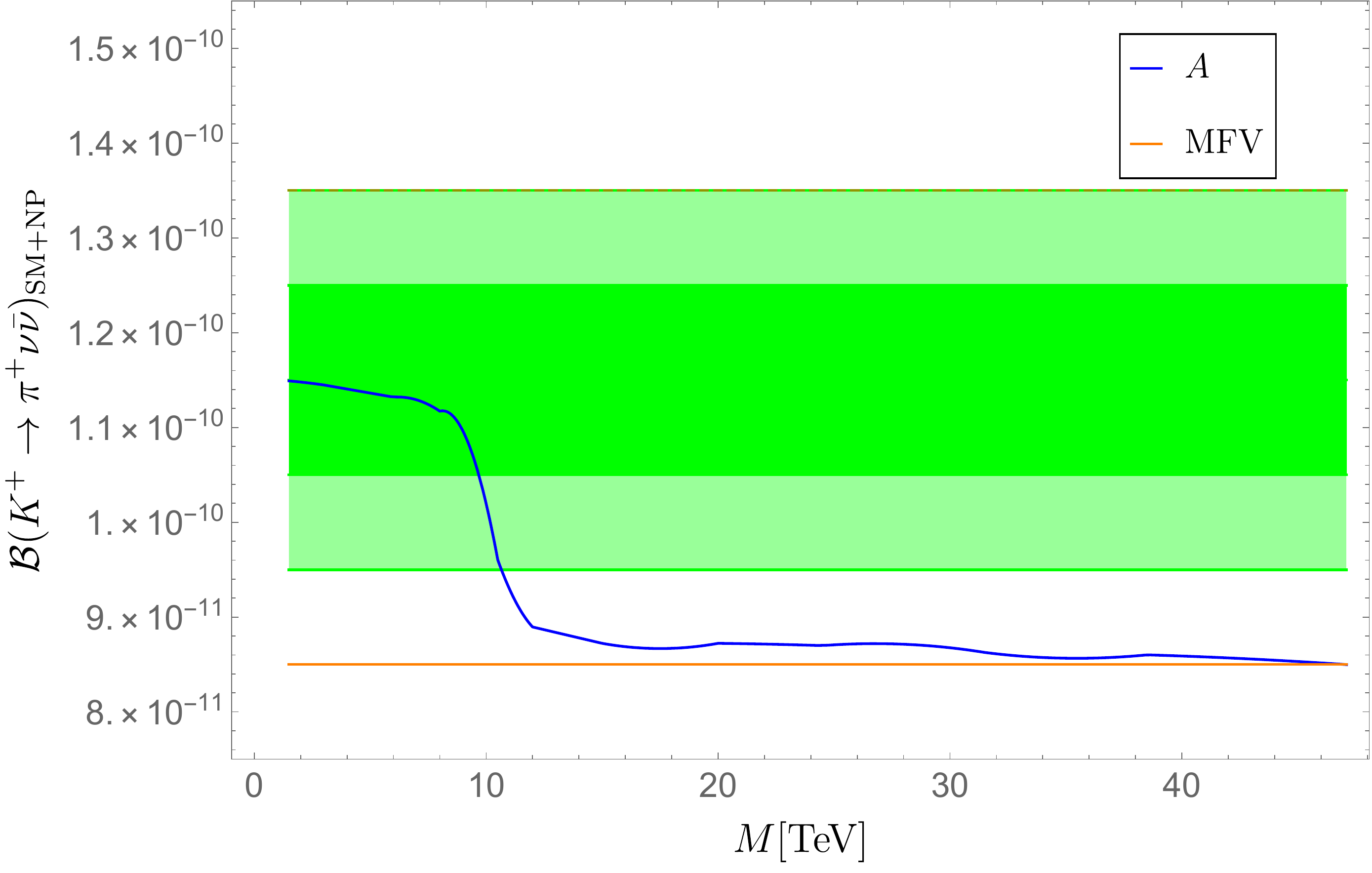}
\caption{Comparison of different flavour models, within the MSSM, in accommodating observables that could exhibit a $3 \sigma$ deviations with respect to the SM, as function of the overall soft-breaking scale $M$ (see main text). The green bands denote the $1 \sigma$ and $2 \sigma$ regions around the central values in the hypothetical future scenario for each observable, while the dashed lines denote the present $1 \sigma$ range. When a given flavour models does not appear on a given plot, it is implied that it yields a completely negligible contribution to the corresponding observable.}
\label{fig:plots}
\end{figure}

\begin{description}
\item[Decoupling.]
With the only exceptions of $|\epsilon_K|$ and $\Im(M^D_{12})$, i.e.~of the $\Delta F=2$ mixing in the $1$--$2$ sector,
for all other observables and for all flavour models, the decoupling limit is reached for $M$ well below $50~{\rm TeV}$.  
On the other hand, in each observable  there is at least one flavour model able to accommodate 
a significant deviation from the SM for $M$ below  $10$~TeV. 

\item[MFV vs. \boldmath{$U(2)$}.]
In these two scenarios  flavour mixing in the MSSM is closely connected to the structure of the CKM matrix.
Interestingly enough, the two frameworks behave quite differently: while the MFV framework is unable to accommodate sizeable effects in any observable,  within the $U(2)$ model large deviations from the SM can occur  in $\Im(M^D_{12})$ and $|\epsilon_K|$. 
This difference can  be understood by the rigidity of the MFV framework, where large off-diagonal entries in the squark mass matrices 
are necessarily accompanied by an overall increase of squark masses. Moreover, in the $U(2)$ model under consideration there is an
effective decoupling of the mixing between third generation and first two generations vs.~the  mixing among the first two generations~\cite{Barbieri:2011ci}. 
This feature allows sizeable effects in $\Im(M^D_{12})$ and $|\epsilon_K|$ that are not restricted by constraints from other observables.

Nevertheless, both in MFV and in the $U(2)$ model the absence of large non-diagonal $A$ terms and the constraints from 
$\Delta B=2$ amplitudes prevent large effects in the  $\Delta F=1$ observables.

\item[\boldmath{$U(1)_{\rm FN}$}.]
As expected, the $U(1)_{\rm FN}$ model is the one which can accommodate large deviations from the SM, even at relatively large $M$ values,
in most of the observables. The only observable  where this does not occur is $\mathcal{B} (K^+ \to \pi^+ \bar{\nu} \nu)$, but this is
because of our particular implementation of the $U(1)_{\rm FN}$ model. As a matter of fact, with non-universal 
$A$ terms as expected in general (e.g. see Eq. \ref{eq:U1_mass_terms}), this framework  can accommodate 
large deviations in  $\mathcal{B} (K^+ \to \pi^+ \bar{\nu} \nu)$ similarly to those shown in the disoriented $A$-term case.

It is worth stressing that in the $|\epsilon_K|$ and 
$\Im(M^D_{12})$ cases the $U(1)_{\rm FN}$ model provides a good fit even for very large values of $M$: the decoupling for $\Im(M^D_{12})$
occurs at $M\sim 300~\rm TeV$ and at even higher energies for $|\epsilon_K|$. As anticipated, 
the price to pay for this `flexibility' is the tuning  of the model parameter at low energies. 

\item[Disoriented \boldmath{$A$} terms.]
This flavour model is the one accommodating the largest effects in the two $\Delta F=1$ observables, at fixed $M$.
This is because of the potentially large impact in down-type $\Delta F=1$ amplitudes induced by  
chargino-mediated $Z$-penguin diagrams.
Interestingly, and somehow unexpectedly, the  disoriented $A$-term framework provides a (slightly) better fit to $\epsilon_K'/\epsilon_K$ with respect to the  $U(1)_{\rm FN}$ one. As in the case of $\mathcal{B} (K^+ \to \pi^+ \bar{\nu} \nu)$ mentioned above, 
this is a consequences of our illustrative choice of neglecting $A$-term contributions in the $U(1)_{\rm FN}$ case. 
 
 The large difference between the tiny effects in $\Delta F=2$ observables vs.~the large
effects in ${\Delta F=1}$ observables provides a distinctive signature of the $A$ terms. This fact is a general consequence of the 
$SU(2)_L$--breaking nature of the $A$ terms, which imply that their contribution in $\Delta F=2$ box diagrams leads to 
effective dimension-8 operators (with four fermion and two Higgs fields), which are strongly suppressed for $M > 1$~TeV.
\end{description}

Beside these specific observations, a key feature emerging from this analysis is the complementary role of these six observables 
in reconstructing the structure of the soft-breaking terms, in the hypothesis that one (or more) of them will exhibit a deviation from the SM in the 
near future.  Figure \ref{fig:plots} shows indeed that each model is associated to a characteristic signature with large or small effects, at a given scale,
in a given subset of observables.

\section{Conclusions}
The flavour structure of the soft-breaking terms  has always been considered one of the most puzzling aspects of the MSSM.
The necessity to raise the overall scale of  SUSY breaking, in order to satisfy the null results of direct 
searches, has partially ameliorated the SUSY  flavour problem: while with soft masses in the $\mathcal{O}(100)$~GeV
regime only MFV-like scenarios were compatible with low-energy data, with a SUSY breaking scale  
of a few TeV (or above) many more options 
are allowed. However, a residual flavour problem remains in the motivated range of soft-masses below $100$~TeV, 
which preserves many 
of the virtues of the MSSM. On the one hand, this motivates 
a non-trivial flavour structure for the soft-breaking terms. On the other hand, this implies that low-energy flavour-changing 
observables might represent a unique opportunity to probe the structure of the MSSM in the next few years. 

In this paper we have provided a quantitative estimate of the sensitivity of selected flavour-physics observables
to four distinct (and motivated) frameworks characterising the flavour structure 
of the soft-breaking terms for soft-masses up to $\mathcal{O}(100)$~TeV.
We have shown that even in scenarios where flavour mixing is CKM-like, specifically in the case of a chiral $U(2)$ flavour mixing,
$\Delta F=2$ observables could exhibit deviations from the SM for  superpartners 
as heavy as $10$~TeV. In this regime the model would escape direct searches in the high-luminosity phase of the LHC,
while the  deviations in low-energy observables would a have realistic chances to be detected
(combining improvements both on the theory and on the experimental side).

More generally, we have shown that a set of six  low-energy observables with realistic
chance of improvements in the near future, namely $\Delta M_{s,d}$,  $\epsilon_K$, 
 $\epsilon_K'/\epsilon_K$, $\mathcal{B} (K\to \pi\nu\bar\nu)$, and the phase of $D$--$\bar D$ mixing,
 could play a very important role in characterising the flavour structure of the MSSM
 with heavy  superpartners. Our findings are summarised by the 
 plots in Figure \ref{fig:plots}, which illustrate the capability of each of the four flavour models
 to fit a realistic deviation from the SM in each of these observables, 
 as a function of the overall SUSY breaking scale. 
 
 One of the most striking features of the six observables in Figure \ref{fig:plots} is their 
 complementarity in the regime of quark masses of $\mathcal{O}(10)$~TeV, which 
 is the regime that could be probed at the FCC-hh (see Fig.~38 in \cite{Golling:2016gvc}).
This provides a clear illustration of the   
importance of  low-energy flavour-changing observables even in a long-term perspective. 
 We stress that this is not a generic statement valid only in models with flavour-anarchic
 soft-breaking terms, which lack theoretical justification. As we have shown with our four 
 explicit examples, it remains true also in very motivated models based on approximate 
 flavour symmetries.   
 
 We finally note that further key low-energy constraints (and possibly hints) on the high-scale
 MSSM could be obtained by lepton-flavour violating observables and by the 
 flavour-conserving electric-dipole moments of quarks and leptons~\cite{Raidal:2008jk,Strategy:2019vxc}.
 For the sake of simplicity, we focused the present analysis only 
 on quark flavour-violating observables, since these are the ones for which the relevance of 
 future high-precision measurement  was less obvious. If the MSSM is the ultraviolet completion of the SM,
 the improvement on all these low-energy observables is a necessary tool to reconstruct its 
 flavour structure from data. 

\section*{Acknowledgements}
 
We thank Dario Buttazzo for useful discussions in the early stage of this work.  We are also grateful to Amogha Pandeshwar for providing advice regarding the parallel programming employed during the multidimensional minimization procedure. This project has received funding from the  European Research Council (ERC) under the European Union's Horizon 2020 research and innovation programme  under grant agreement 833280 (FLAY), 
and by the Swiss National Science Foundation (SNF) under contract 200021-159720.

\end{document}